\documentclass[onecolumn,%
 aip,
rsi,%
 amsmath,amssymb,
 reprint,%
]{revtex4-1}

\usepackage{graphicx}
\usepackage{dcolumn}
\usepackage{bm}
\usepackage {CJK}
\begin{document}
\begin {CJK*} {GB} {}
\title{Toward the analogue of a thermally generated electromagnetic field}
\author{Bob Osano} 
\email{bob.osano@uct.ac.za} \affiliation{Cosmology and Gravity Group, Department of Mathematics and Applied Mathematics, University of Cape Town, Rondebosch 7701, Cape Town, South Africa}
\affiliation{Academic Development Programme, Science, Centre for Higher Education Development, University of Cape Town, Rondebosch 7701, Cape Town, South Africa}
\author{Patrick W. M. Adams}
\email{patrick.adams@uct.ac.za} \affiliation{Cosmology and Gravity Group, Department of Mathematics and Applied Mathematics, University of Cape Town, Rondebosch 7701, Cape Town, South Africa}

\begin{abstract}
The evolution equations for magnetic and vorticity fields are known to display the same pattern when dissipation and sources terms are neglected. We investigate the analogy between the two fields for non-vanishing dissipation and sources. In addition to the magnetic Reynolds (${Re_{M}}$) and Prandtl (${Pr}_{M}$) numbers, we define a new number (${ S}_{M}$) that is given by the ratio of the diffusive term to the Biermann battery term and which allows for a different classification of magnetized fluid behavior. Numerical simulations of the two fields are then carried out given a parameter space made of Reynolds, Prandtl and source numbers. We find it appropriate to present and discuss the findings against Prandtl numbers given that these provide the link between viscous and magnetic diffusion. Our simulations indicate that there exists a range of Prandtl numbers for which the fields remain analogues which raises the important question of how far the analogy goes, and also raises the prospect of a theory of analogue magnetism. 
\end{abstract}

\date{\today}
\maketitle
\end{CJK*}
\section{Introduction} 
Magnetic fields seem to exist everywhere in the universe. The need to explain their presence has led to a number of theories related (i) to their formation (magneto-genesis) or (ii) to their growth (dynamo). One magneto-genesis theory suggests that there were mechanisms that existed that could generate cosmic magnetic fields after rather than in the Big Bang. Efforts to identify these mechanisms have remained sustained, as is seen in research literature, although an effective mechanism remains elusive. An incomplete list of such efforts may be found in the following articles, review-articles on cosmic magnetic fields \cite{Bier, Rev, Rev1, Rev3} and in the articles cited in them. Such mechanisms could have operated after recombination during which seed fields of magnitude $B\sim10^{-20}$ G may have been generated. The Biermann battery\cite{Bier}, which leaves a non-vanishing $\nabla T_{e}\times \nabla n_{e}$ (where $\nabla T_{e}$ is the gradient of the temperature of electrons and $\nabla n_{e}$  is the gradient of the electron number density) is an example of a mechanism that can generated magnetic fields given the correct conditions. 

\par In the context of the universe, the battery mechanism can operate during the period of galaxy formation and can generate such fields \cite{Kulsrud,Cowling}. The fields generated via this mechanism are small in magnitude and therefore there exists a disjuncture between what is observed today which of $B\sim10^{-6}$G in magnitude, and what should be observed if no other mechanism operated on the seed field between the time they were generated and today. This is obviously simplistic given our current understanding of the possible evolution of the universe, but therein lies the problem. How did these fields become so strong? Some amplification mechanisms must act on the seed fields if they are to grow. It seems that different amplification mechanisms, each with a different threshold, are necessary to achieve the present day magnitudes \cite{Kulsrud}. Magnetic fields have remained difficult to understand firstly because of the technical difficulties encountered in the efforts to observe them, which arise mainly from the fact that these fields are weak and because of the large distances between the source and the observer. On the other hand the physics of plasma on galactic scales is also not well understood, which accounts for some of the discrepancies between observation and theoretical predictions. These sets of challenges call for a different approach.\\

\par We think that  {\it magnetic analogues} are a viable option. Analogues provide an indirect way of studying phenomena that may not readily lend themselves to physical experimentation. A suitable analogue can therefore be a useful tool in providing insight into an otherwise intractable problem. But a complete theory of magnetic analogues  is not yet available, although bits and pieces that could make up the theory are strewn in relevant scientific literature. In this article, we examine a correspondence between the magnetic field and fluid dynamics that goes deeper than what is presently available, with the aim of contributing to the searching for a theory of magnetic field analogue. As it has been pointed out elsewhere \cite{LIV1}, analogy is not identity. This means that one cannot say that {\it analogues} are completely equivalent to magnetic fields. Nevertheless such models would be expected to accurately show the key features of magnetic fields. In two articles \cite{BP, PB} , we examined the analogues of magnetism where the battery terms were necessarily turned off, but not the dissipation. Earlier work on the search for analogues is found in \cite{Ela1, Ela2, Ela3, Ela4, Bat1, Cowling, Park}, where hydrodynamic dynamo models were considered. In this article, we go beyond hydrodynamics and consider fluids with $\nabla p\neq 0$, with the intention of examining the Biermann battery term effect.
\section{Relevant analogous equations}
There are two separate systems of equation that concern us: (i) that of a charged fluid ( which encode the growth behavior of the magnetic flux), and (ii) that of non-charged viscous fluid ( as given by the vorticity field). The second case is nuanced and deserves further clarification. If one considers a weakly ionized fluid, like we do in the rest of the article, the fluid mixture is made up of electrons, ions and neutral particles. In this scenario, the electron-ion fluid naturally interacts with the neutral particles (assuming a constant ionization fraction) via particle collisions frictions leading to ambipolar diffusion. 

This opens up the possibility of comparing the magnetic flux not just with the vorticity of the corresponding magnetic fluid but with the vorticity of any general viscous fluid. Herein lies the potential for a theory of analogue magnetism. With this in mind, we devote remainder of this section to a discussion of the main systems of equations governing these flows.
\subsection{Magnetic induction equation}
Our focus is on the modified induction equation, modified in the sense that it carries a source term.
\\\begin{eqnarray}
\label{eq:1}\frac{\partial{\bold{B}}}{\partial{t}}-\nabla\times\bold{v}\times\bold{B}-\eta\nabla^{2}\bold{B} &=&\frac{c\nabla p_{e}\times\nabla n_{e}}{n_{e}^2 e(1+\chi)},
\end{eqnarray} where $\bold{B}$ is the magnetic flux, while $\eta$  is the coefficient of diffusion. Subscripts $e$ stands for electrons, while $p$ and $n$ are pressure and number density respectively.
As pointed out in \cite{Axel, Kulsrud}, if one were to assume that the charged fluid is weakly ionized and therefore made up of three parts: free electrons, protons and hydrogen atoms and where the ionization fraction (constant in space, for a given temperature) is denoted by $\chi$, it is easy to recast the equation with the setting $p_{e}=\chi p_{_{(cf)}}/(1+\chi)$, $n_{e}=\chi\rho_{_{(cf)}}/m_{p}$ and ${\mathcal{B}}=e(1+\chi)B/m_{p}$ (other possible parameterizations exist, see for example \cite{Bob}). The subscript ({\it cf}) stands charged fluid. $m_{p}$ is the proton mass. This conversion yields:
\begin{eqnarray}
\label{eq:1ba}\frac{\partial{\bold{\mathcal{B}}}}{\partial{t}}-\nabla\times\bold{v}\times{\mathcal{B}}-\eta\nabla^{2}{\mathcal{B}} &=&\frac{c\nabla p_{_{(cf)}} \times\nabla\rho_{_{(cf)}}}{{{\rho^{2}}_{\small{_{(cf)}}}} },
\end{eqnarray} the form that we will use in the rest of the analysis. 
\subsection{Fluid vorticity}
In this section we briefly review fluid vorticity. It is important to emphasize that the vorticity that we consider in this section is not the magnetic fluid vorticity, but that of a neutrally charged fluid. The idea is to compare and contrast the behavior of the magnetic fluid to a non-charged fluid for the sake of identifying an analogue to the magnetic fluid. Let ${\bf u}$ be a vector field describing the motion of a neutrally charged fluid at a given temperature. Such a fluid has a vorticity vector $\omega=\nabla\times{\bf u}$, which is necessarily solenoidal ($\nabla {\bf .} \omega=0$). It is relatively easy to show that the vorticity vector obeys the propagation equation:
\begin{eqnarray}
\label{eq:2}\frac{\partial{\bold{\omega}}}{\partial{t}}-\nabla\times\bold{u}\times\bold{\omega}-\nu\nabla^{2}\bold{\omega} &=& - \frac{\nabla p_{_{(nf)}} \times\nabla\rho_{_{(nf)}}}{{{{\rho}^{2}}_{_{(nf)}}} }.
\end{eqnarray}
We note that the source terms are comparable if one sets  the speed of light to unit ($c=1).$ The subscript ${\it nf}$ in Eqn (\ref{eq:2}) indicates that the fluid is neutral in this case. We use this notation to distinguish it from the charged fluid ($cf$) given in the previous section. Vorticity is vital to the understanding of how fluids flow. It will be noted that its importance stems from the fact that one can recover fluid velocity field by inverting $\omega=\nabla\times{\bf u}$ to find an integral over  the velocity field, although this is subject to conditions \cite{Vort}. If the viscosity is negligible, the fluid will have a barotropic equation of state {assuming no other source of heating exists and a uniform temperature}, and the external forces will be conservative and the vorticity will satisfies Helmholtz's laws. These laws necessarily imply that vorticity can not be created and hence the Helmholtz equation:
\begin{eqnarray}
\label{eq:2d}\frac{\partial{\bold{\omega}}}{\partial{t}}&=&\nabla\times(\bold{u}\times\bold{\omega})=-\bold{u}\bold{.}\nabla\bold{\omega}+\bold{\omega}\bold{.}\nabla\bold{u} -\bold{\omega}\nabla\bold{.}\bold{u}.
\end{eqnarray} It follows that a violation of Helmholtz's laws is necessary for the generation of vorticity. To this end, we consider non-negligible and hence non-barotropic equation of state in our analysis. We also set all external forces to zero for simplicity, where not even the gravitational force is allowed ( the alternative will be considered in \cite{Bobgrav}). We now consider equations (\ref{eq:1ba}) and (\ref{eq:2} ) together.
\subsection{The analogous system\label{units}}
We investigated and confirmed the well known analogy between magnetic field, ${\bold B}$, and vorticity field, $\omega$, equations in  \cite{BP} , wherein we also extended the work to the special case where the magnitude of diffusion equals that of kinetic viscosity. This work has further been extended in \cite{PB} where simulations and comparisons of various fluids types have been done. It will be noted that the studies just mentioned looked at systems of equations that did not have any source terms. The present study examines the analogy between magnetic field and vorticity field equations in cases where the sources terms are non-negligible and in particular where the Biermann's term is non-vanishing. 
The relevant comparative equations in this case take the form:
\begin{eqnarray}
\label{eq:1b}\frac{\partial{\bold{\mathcal{B}}}}{\partial{t}}-\nabla\times(\bold{v}\times{\mathcal{B}})-\eta\nabla^{2}{\mathcal{B}} &=&\frac{\nabla p_{_{(cf)}} \times\nabla\rho_{_{(cf)}}}{{{\rho^{2}}_{\small{_{(cf)}}}} },\\
\label{eq:2b}\frac{\partial{\bold{\omega}}}{\partial{t}}-\nabla\times(\bold{u}\times\bold{\omega})-\nu\nabla^{2}\bold{\omega} &= &- \frac{\nabla p_{_{(nf)}} \times\nabla\rho_{_{(nf)}}}{{{\rho}^{2}_{_{(nf)}}}},
\end{eqnarray} where $c=1$. Each of these equations couple to their respective Navier-Stokes equation:
\begin{eqnarray}
\label{ns101}\frac{\partial{\bold{v}}}{\partial{t}}+\bold{v}\bold{.}\nabla{\bold{v}}-\nu\nabla^2\bold{v}&=&-\frac{\nabla p_{_{(cf)}}}{\rho_{_{(cf)}}}\\
\label{ns102}\frac{\partial{\bold{u}}}{\partial{t}}+\bold{u}\bold{.}\nabla{\bold{u}}-\nu\nabla^2\bold{u}&=&-\frac{\nabla p_{_{(nf)}}}{\rho_{_{(nf)}}}.
\end{eqnarray} It would appear, from equations (\ref{eq:1b} and \ref{eq:2b}) that $\omega$ should ideally be compared to $-\mathcal{B}$ which, in practical terms, is a matter of the orientation of the field. For example, it has been noted in \cite{Krus} that in some 4 out of 5 galaxies, the radial component of the spiral field could be such that the field points inward. This suggests that the induction equation and the related nonlinearities do not distinguish between {\bf B} and {\bf - B} other than in the case of Hall effect where direction is important \cite{Holl}. In any case, the $\bf{B}_{rms}$

We proceed to simulate and compare various sub-cases of the two systems. Simulations are done using the \textsc{Pencil Code} \cite{Pen} for a periodic of domain $2\pi\times2\pi\times2\pi$ and dimensions $32^3$.\setlength{\parskip}{0mm}
The calculations in PENCIL CODE are unit-agnostic, in the sense that all results remain the same independent
of the unit system in which one interprets the numbers.  For example, if one simulate
a simple hydrodynamical flow in a box of length $L = 1$ and finds a maximum velocity $V_{Max} =0.5$ after $t = 3$ time units, then one may interpret this as $L = 1m$, $V_{Max} = 0.5m/s$,
$t = 3 s$, or as $L = 1 pc$, $V_{Max} = 0.5 pc/Myr$, $t = 3Myr$, depending on the physical system one has in mind. The units one uses must nevertheless be consistent. This means that in the second example above, the units for diffusivities would be $pc^{2}/Myr,$ etc. The Reynold numbers, to be considered later in the article, may also be interpreted in this way. For example, if viscosity $\nu$ and magnetic diffusivity $\eta$ are set to $5\times10^{-3}$, $V_{Max}=0.3$, the mesh Reynolds number is about $V_{rms}\times\delta x/\nu = 0.3\times(2\pi/32)/5\times10^{-3}\approx 12$ (see \cite{Pen} for full explanation). We will define relevant Reynolds numbers later in the article.
\section{Simulation results and discussion}
There are four different comparisons that we examine in this study. (i) The first case compare the evolution of magnetic flux when the battery term is present to the case where it is absent. (ii) Similar, but separate, comparison is done for vorticity field equations of a non charged fluid. (iii) We then compare that magnetic ($cf$)and vorticity($nf$) field equations with battery term, and (iv) without the battery term. These simulations allow us to examine the effect of pressure on the general structure of the magnetic and vorticity fields.  It is instructive to note that in the case where pressure is present, pressure terms appear in both the magnetic and vorticity field equations and the respective Navier - Stokes equations to which these field equations are coupled as shown in equations(\ref{eq:1b} -\ref{ns102}). We first examine each field individually. \setlength{\parskip}{0mm}
\subsection{Evolution of magnetic field}
The magnetic induction (hereafter $\mathcal{MI}$) and the Navier - Stokes (hereafter $\mathcal{NS}_{cf}$) equations are coupled, with the level or the strength of coupling heavy depended on the kind of forces at work. For simplicity we choose to neglect the Lorentz and all other external forces, wherein we are interested in the comparative analysis. The Lorentz force is crucial in nonlinear dynamos. It is known that presence of this force causes the dynamo to stop growing because this force changes the flow in such away that the dynamo action is reduced leading to dynamo saturation, but this is not the subject of this article. With vanishing Lorentz force, the coupled system takes the form:
\begin{eqnarray}
\label{one1}\frac{\partial{\bold{\mathcal{B}}}}{\partial{t}}-\nabla\times(\bold{v}\times{\mathcal{B}})-\eta\nabla^{2}{\mathcal{B}} &=&\frac{c\nabla p_{_{(cf)}} \times\nabla\rho_{_{(cf)}}}{{{\rho^{2}}_{\small{_{(cf)}}}} },\\
\label{ns1}\frac{\partial{\bold{v}}}{\partial{t}}+\bold{v}\bold{.}\nabla{\bold{v}}-\nu\nabla^2\bold{v}&=&-\frac{\nabla p_{(cf)}}{\rho_{(cf)}}.
\end{eqnarray} We also use a modified set of this system where pressure has been dropped ($p=0$) in both equations and wherefore in order to keep track of which system is being simulated, we use the notation $\mathcal{B}_{123}$ to denote $\mathcal{B}$ in equation (\ref{one1}), indicating that the magnetic field experience the effects of the three terms i.e.    
\begin{eqnarray}
\label{one1a}\frac{\partial{\bold{\mathcal{B}}}}{\partial{t}}&=&\underbrace{\nabla\times(\bold{v}\times{\mathcal{B}})}+\underbrace{\eta\nabla^{2}{\mathcal{B}}} +\underbrace{\frac{c\nabla p_{_{(cf)}} \times\nabla\rho_{_{(cf)}}}{{{\rho^{2}}_{\small{_{(cf)}}}} }},\nonumber\\
&&~~~~~~~(1)~~~~~~~~~~~(2)~~~~~~~~~~~~~~(3),\end{eqnarray} where the numbers are used to label the different terms.
$\mathcal{B}_{12}$ to denote the counterpart of this equation that does not have the pressure term (the pressure term is also set to zero in the corresponding $\mathcal{NS}$). 
We inject an initial seed field of strength $10^{-5}$ G for the purposes of kickstarting the process. We monitor the {\it rms} of both $\mathcal{B}_{123}$  and $\mathcal{B}_{12}$. We also set $\eta=\nu=10^{-10}$ for illustrative purposes (very small dissipation in $\mathcal{MI}$ and its corresponding $\mathcal{NS}_{cf}$). The seed field and dissipation values are deliberately chosen to allow for a growing magnetic flux. It is obvious that a strong dissipation will lead to a decaying field and hence not allow for an effective analysis of the contribution of the battery term. The results of the simulations are presented in Fig. (\ref{fig:gull1}).\\
 
 \par In order to understand the structure displayed in Fig. (\ref{fig:gull1}), a closer examination of the two systems given by equations ( \ref{one1}) and (\ref{ns1}), and their pressure-free counterparts, is required. The two growth curves have two points of intersection. The first point of intersection represents the initial equality when only the seed field is present. The seed field then experiences both the amplification due to the dynamo term and the dissipation due to the diffusivity term present. The effect of these two on the magnetic field will differ depending on whether or not the battery term is present. We first consider $\mathcal{B}_{12}$ for a plasma fluid by considering the magnetic Reynolds numbers which allow us to compare the relative effects of the two terms on the strength of the magnetic fields. We note that $Re_{M}=1$ indicates equality and therefore cancelation of the effects. $Re_{M}\ll 1$ and $Re_{M}\gg1$ signify the perfectly conducting limit (frozen-in flux) and the diffusion equations respectively. 
 
 \par So what happens when a source term is present?  In this case it is helpful to define a new dimensionless quantity ${S}_{M}$ given by the ratio of the source term to the diffusive term (The reader is referred the appendix (\ref{appBat}) for a dimensional analysis of the this term). The growth pattern of the field can then be assessed based on both the magnetic Reynolds number and the Source number. We now show that in the presence of battery term growth can still be experienced even if $Re_{M}<1$ for the case where a source term exists. For this to occur, the sum of the magnetic Reynolds number for an induction equation that possess a battery term and the Source number should be greater than one ($Re_{M}$ +${S}_{M}) >1$. The evidence for this is found in the simulation results that we will discuss shortly. It is obvious therefore that ($Re_{M}$ +$S_{M}) \gg1$ will lead to a battery-aided dynamo, which is an amendment to the standard dynamo theory. Although these two numbers do allow us to examine the dynamo theory in the context of battery term, the understanding of the actual dynamics requires one to examine the coupling of the induction equation to the {Navier-Stokes} equation.
  \\\\
 \par In particular, it would appear that the presence of pressure reduces the effect of the dynamo on the field, which can be explained given its effect on the $\mathcal{NS}_{cf}$ equation(\ref{ns1}). This reduction is however countered by the contribution from the battery term that increases the strength over and above what was lost via a reduced dynamo effect. The field with the battery term will therefore momentarily have a higher value than the field without. But while the dynamo effect grows because of the change in velocity via the $\mathcal{NS}_{cf}$, the battery term remains constant in time. This means that over time, a point is reached where the effect of the battery term is balanced by the increased dynamo effect and the diffusion which leads to the second point of intersection. From this point on, the reduced effect of dynamo due to the pressure term is much bigger than what can be compensated by the battery term and hence $\mathcal{B}_{12}$ will grow much faster than $\mathcal{B}_{123}$. It might be interesting to determine how the presence of a battery term modifies the Backus conditions\cite{Back} necessary for a dynamo, something that will be pursued in future.
\subsection{Evolution of vorticity} 
As in the case of $\mathcal{MI}$ and $\mathcal{NS}_{cf}$, we consider the evolution of the vorticity equation (hereafter $\mathcal{VI}$) and its corresponding Navier - Stokes equation (hereafter $\mathcal{NS}_{nf}$). 
\begin{eqnarray}
\label{oneomega}\frac{\partial{\bold{\mathcal{\omega}}}}{\partial{t}}-\nabla\times(\bold{u}\times{\mathcal{\omega}})-\eta\nabla^{2}{\mathcal{\omega}} &=&-\frac{\nabla p_{_{(cf)}} \times\nabla\rho_{_{(nf)}}}{{{\rho^{2}}_{\small{_{(nf)}}}} },\\
\label{NS11}\frac{\partial{\bold{u}}}{\partial{t}}+\bold{u}\bold{.}\nabla{\bold{u}}-\nu\nabla^2\bold{u}&=&-\frac{\nabla p_{(nf)}}{\rho_{(nf)}}.
\end{eqnarray} Here too we use the notation $\omega_{123}$ for equation (\ref{oneomega}) and $\omega_{12}$ for the case without the pressure term. The results of the simulations are given in Fig. (\ref{fig:tiger1}). The graphs are manifestly different because of the pressure term. We also define a fluid Reynolds number ($Re_{V}$) and a Source number $\mathcal{S}_{V}$, the subscript $V$ indicates that these parameters are for the vorticity field system of equations. In this case growth results if $(Re_{V}-\mathcal{S}_{V})>1$
\par As in the case of magnetic field, we find two points of intersections. The first point is the initial equality when only the seed field is present, the second is where the contributions from the various terms balance out leaving the field with a strength similar to that case where the pressure is absent. Although both curves are decaying beyond the second point of intersection, $\omega_{12}$ remains stronger than $\omega_{123}$ for the same reasons given in the case of magnetic field analysis. In particular, from the $\mathcal{NS}_{nf}$ (\ref{NS11}) the pressure leads to a comparatively lower velocity which in turn leads to decreased amplification effect in equation (\ref{oneomega}), this coupled with the reduction on $\omega_{123}$ due to the source term ensures that the strength lags that of $\omega_{12}$ beyond the second point of intersection. We again warn that the simulations are only done for weak vorticity and the case of stronger vorticity will be given elsewhere \cite{PB}. 

\subsection{Magnetic and Vorticity Induction equations with ${Pr}_{M}=1$ and ${Pr}_{M}\neq1$ }
\begin{figure}
        \includegraphics[width=0.75\textwidth]{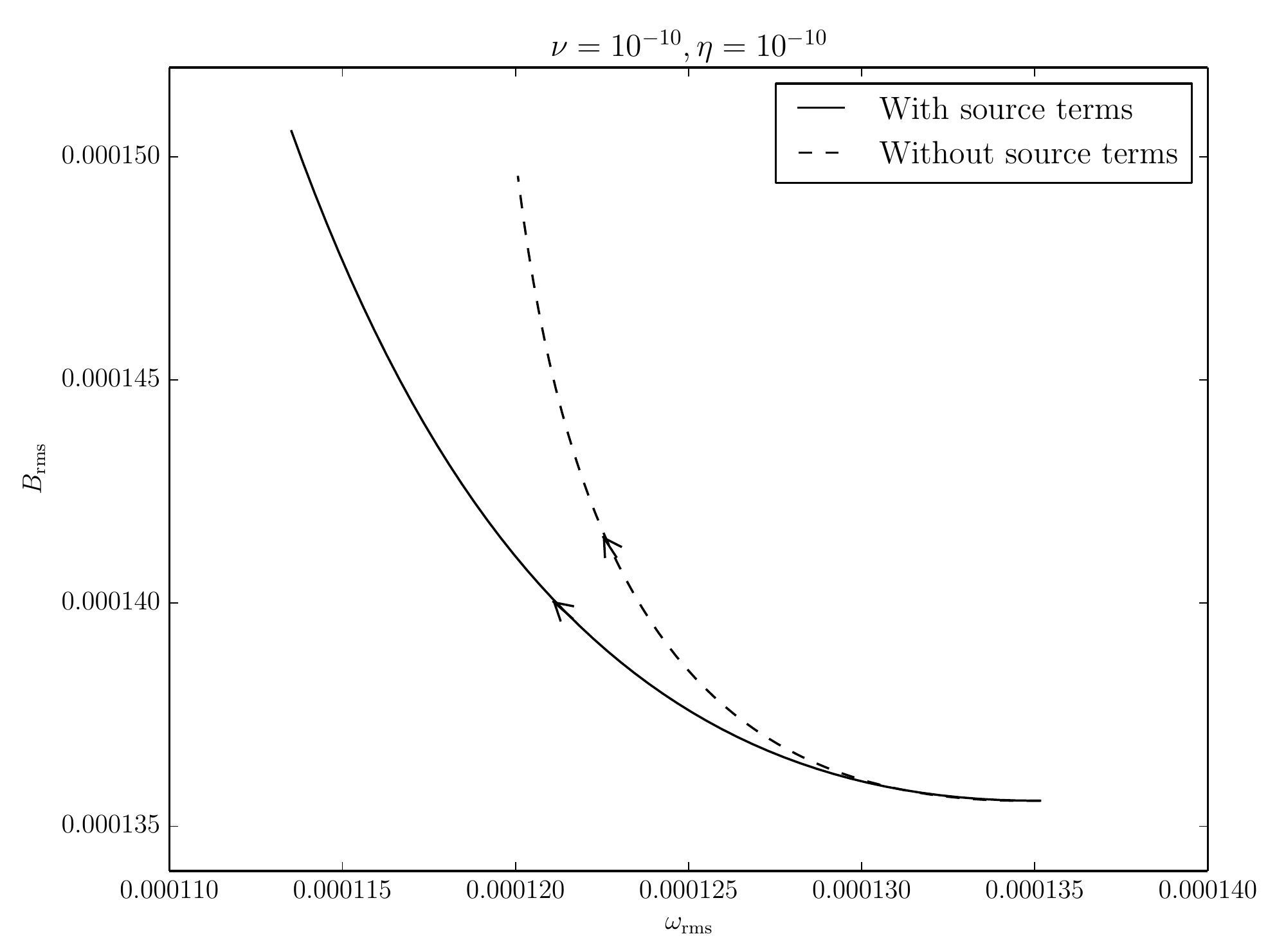}
         \caption{ Again the solid line represents $\mathcal{B}_{123}$ (full equation)  while the dash line represents $\mathcal{B}_{12}$ (without battery term).}
           \label{fig:mouse11}
\end{figure}

\begin{figure}
        \includegraphics[width=0.75\textwidth]{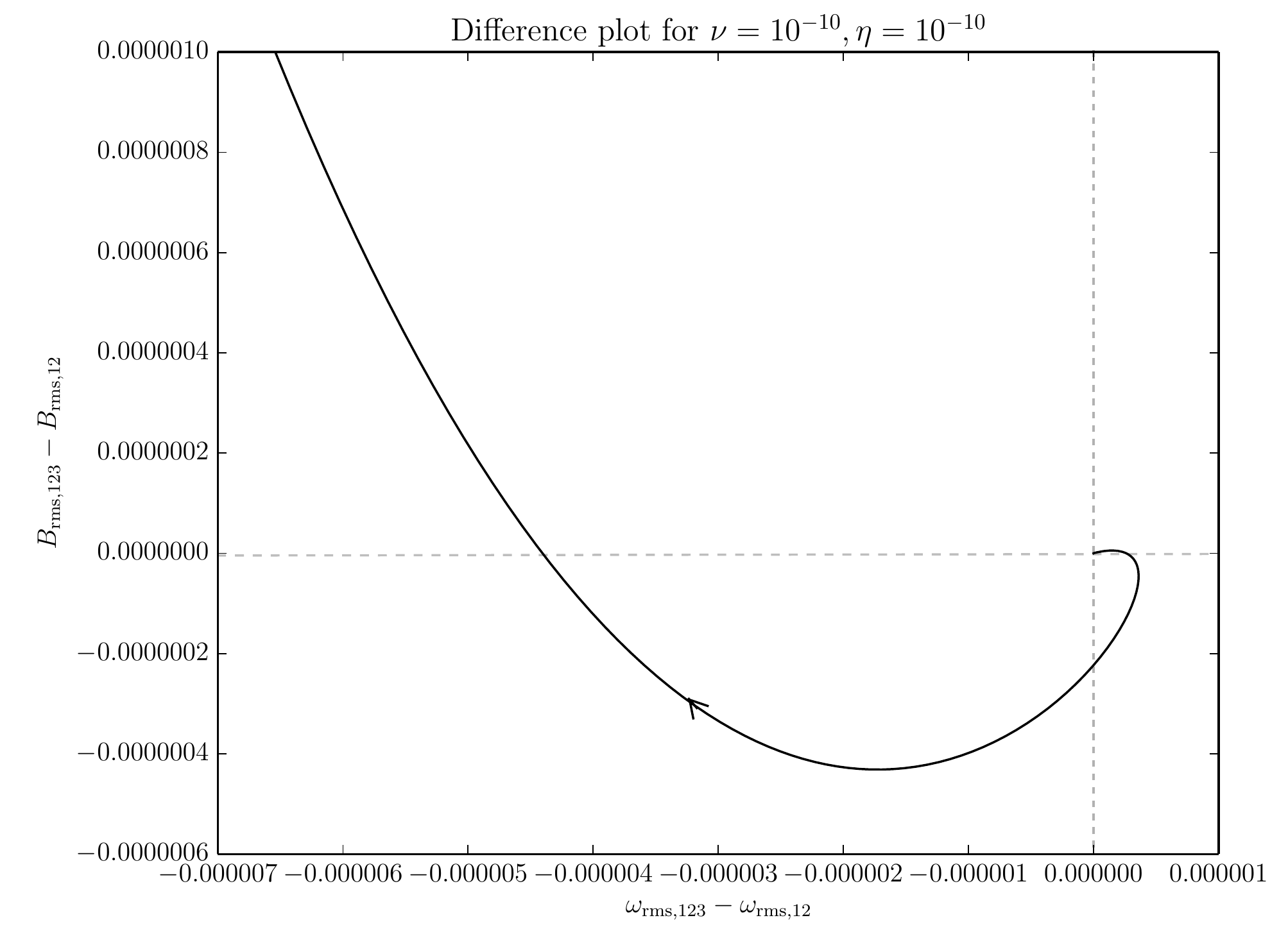}
        \caption{{The graph represents the plot of the time varying difference $\mathcal{B}_{123}-\mathcal{B}_{12}$ against the time varying difference $\omega_{123}-\omega_{12}$}}
      \label{fig:mouse12}
\end{figure}

We are now at a point where we can compare $\mathcal{MI}$ to $\mathcal{VI}$. In order to aid discuss, we use the notation ${Pr}_{M, eff}$ the Prandtl number, where the subscript $eff$ indicates that the ratio  is between magnetic diffusivity and the kinetic viscosity of a general viscous fluid. Two separate cases; ${Pr}_{M,eff}=1$ ( $\nu=\eta=10^{-5}$) and ${Pr}_{M,eff}\neq1$, will be examined for illustrative purposes. These systems encode nonlinear dynamics which often times are too complex to study concurrently. For this reason, we will limit the study to special cases and leave extensions and general cases for future studies. It is sufficient to say that nonlinear effects are important in magnetohydrodynamics because they have the capability to induce dynamo action, where magnetic field lines would be stretched by the flow velocity gradients when this exceeds diffusion. Other nonlinear effects induced by the interaction of a charged  with its environment may be analyzed using the a novel generalization of Maxwell's equations and the resulting induction equation presented in \cite{Bob}.  In the present case however, simulation results are given in Figs. (\ref{fig:mouse11}) and (\ref{fig:mouse12}). Fig.(\ref{fig:mouse11}). 

\begin{figure}
        \includegraphics[width=0.75\textwidth]{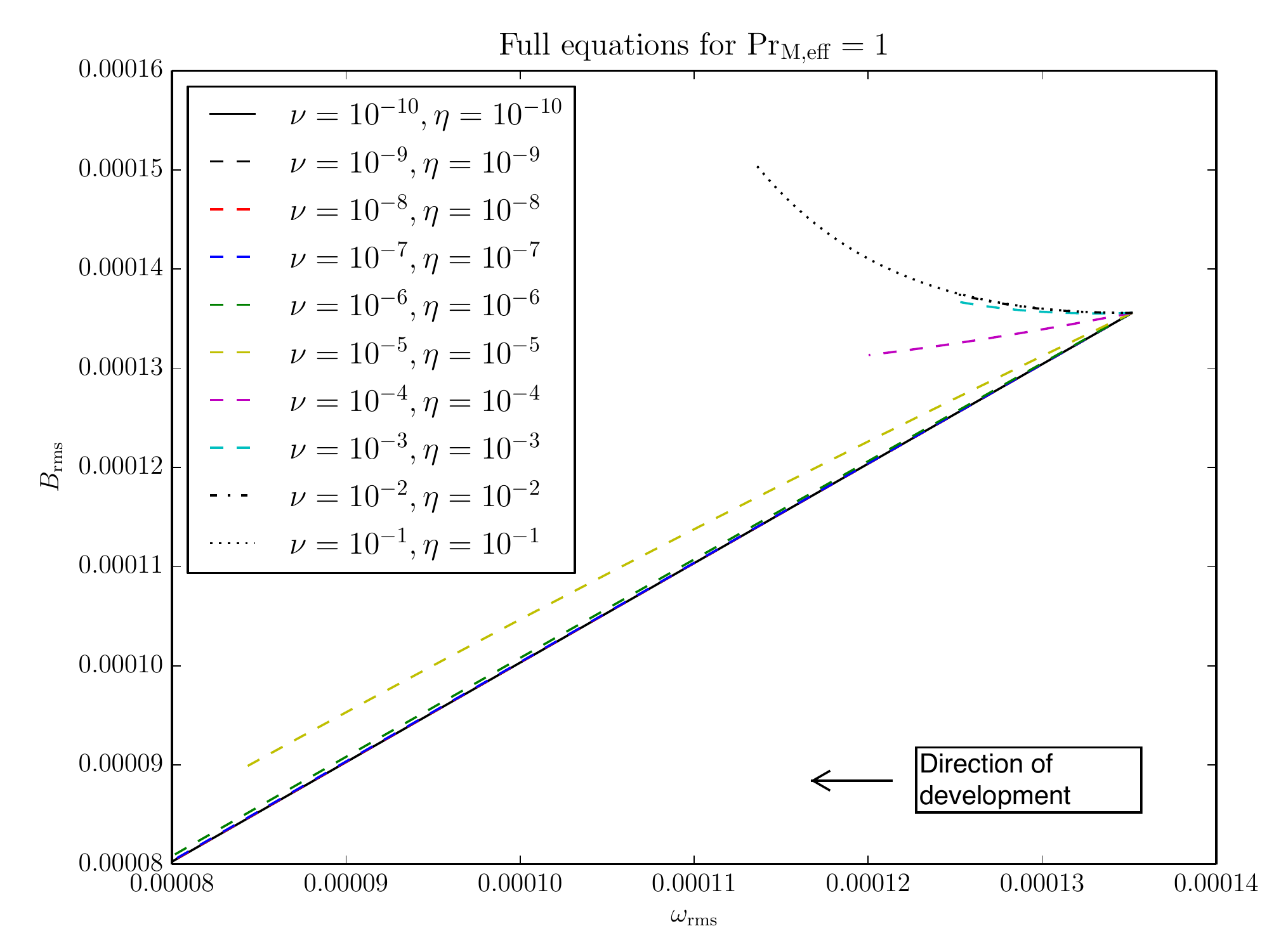}
        \caption{{We have plotted $\mathcal{B}_{123}$ against $\omega_{123}$ for different values of $\nu$ and $\eta$ while keeping ${Pr}_{M}=1$. It is apparent that as dissipation becomes stronger the graph tends toward a straight line. The significance of this diagram is that it indicates that a linear relationship exists for weak non-zero dissipation implying analogy even when dissipation is non zero. }}
      \label{fig:mouse111}
\end{figure}

The curves are best understood by comparing them to a straight line with a positive slope, which would be the case were the two to be analogous. We can therefore conclude that given these values for dissipation, $\mathcal{B}_{123}$ approximates a linear function (albeit poorly) better than $\mathcal{B}_{12}$ does. In fact, we see in Fig. (\ref{fig:mouse111}) that when we make the dissipation stronger the relationship move progressively towards a straight line and hence better analogue. We have only presented the case for $\mathcal{B}_{123}$, given that the case for $\mathcal{B}_{12}$ will show the same structure although with a much slower progression toward as a straight line. What if these structures are the results of systematic effect in the code and have nothing to with the physical effect with which we are associating them with? The way to check this is to obtain a plot of the difference between $\mathcal{B}_{123}$ and $\mathcal{B}_{12}$ against the difference between $\omega_{123}$  and $\omega_{12}$. The idea is that each run will experience the same systematic effects and hence taking the difference eliminates such effects. The results of these simulations are given in Fig.(\ref{fig:mouse12}). \par
\begin{figure}
  		\label{fig:1}
        \includegraphics[width=0.75\textwidth]{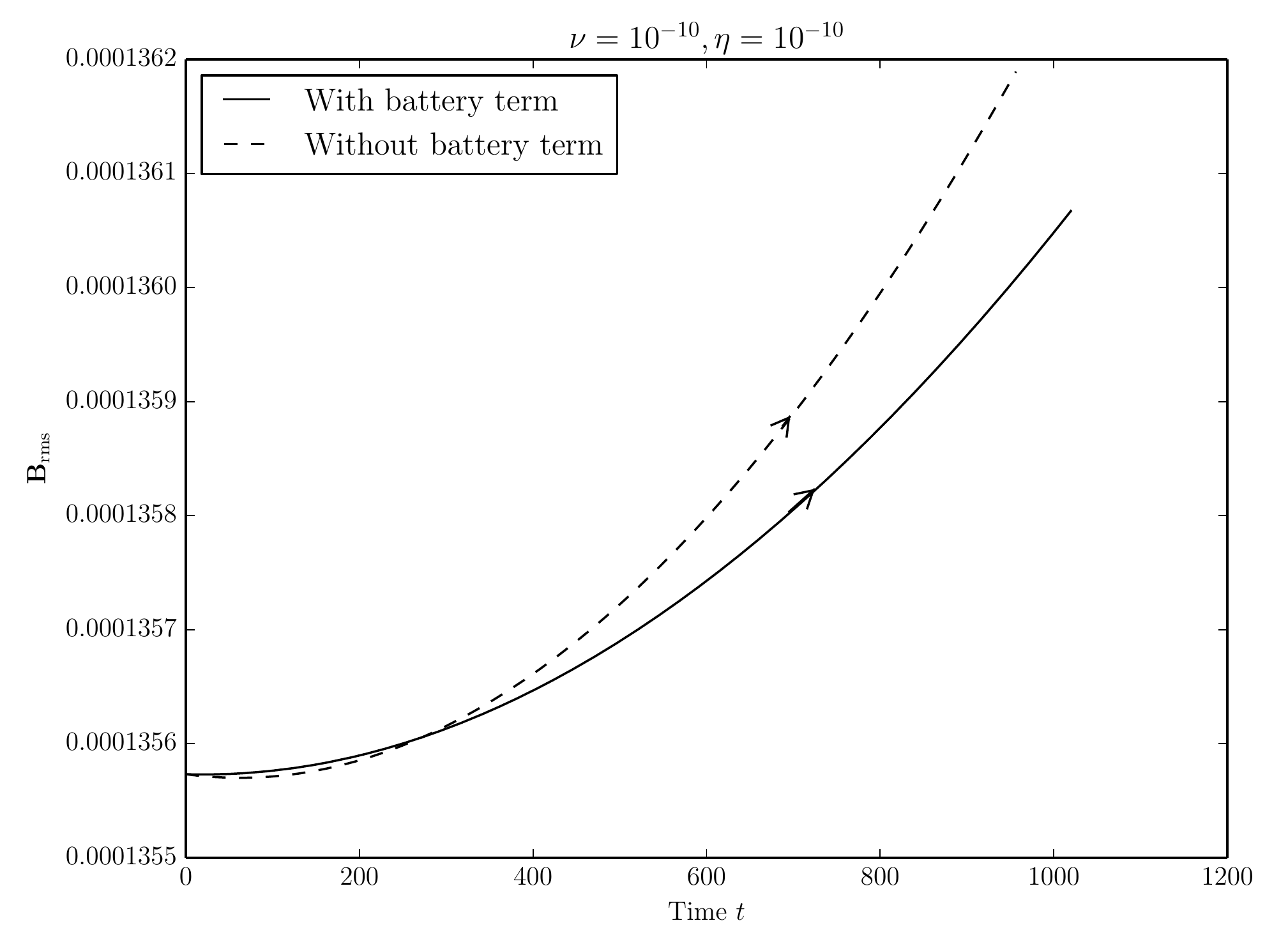}
        \caption{{{The dotted line represents $\mathcal{B}_{123}$ (with battery term or the full equation)  while the solid line represents $\mathcal{B}_{12}.$ (without battery term)}}}
        \label{fig:gull1}
 \end{figure}

\begin{figure}
          \includegraphics[width=0.75\textwidth]{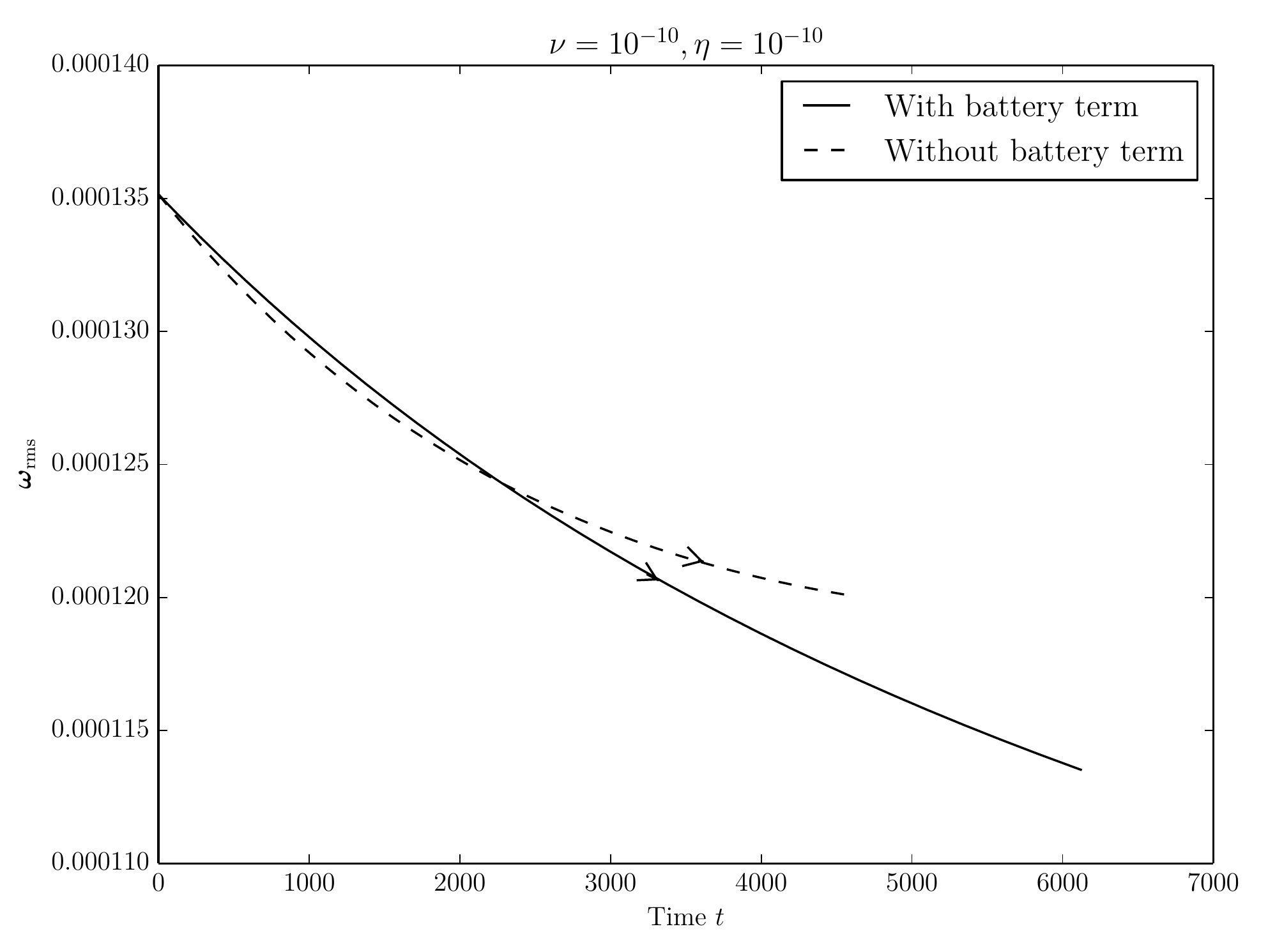}
        \caption{\small{The dotted line represents $\omega_{123}$ (with battery term)  while the solid line represents $\omega_{12}.$ (without battery term)}}
        \label{fig:tiger1}
\end{figure}

\begin{figure}
        \includegraphics[width=0.75\textwidth]{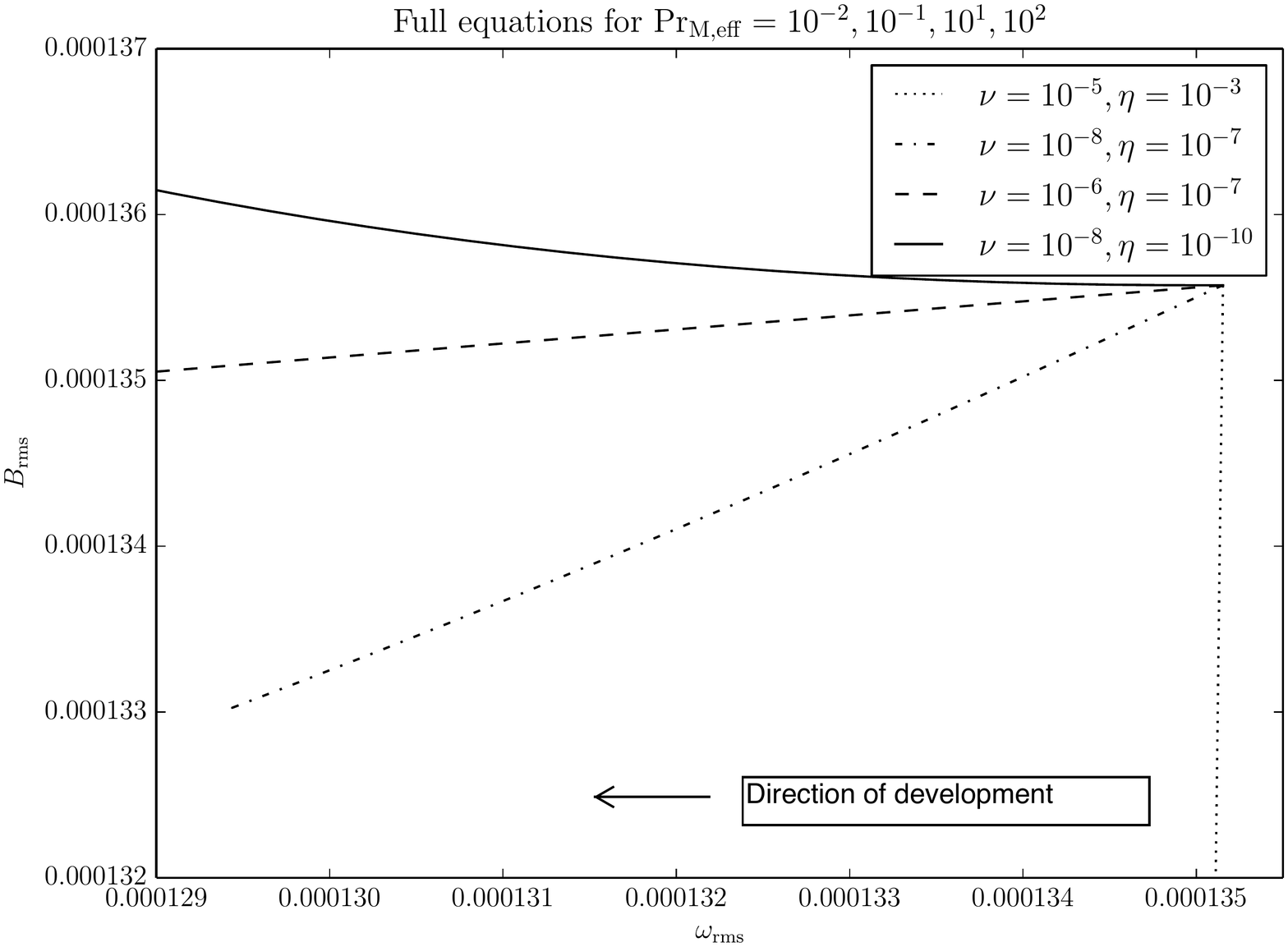}
        \caption{We have plotted $\mathcal{B}_{123}$ against $\omega_{123}$ for different values of $\nu$ and $\eta$ while keeping ${ Pr}_{M}\neq1$. This diagram shows that the Prandtl number need be comparatively for an analogue to be viable. In particular, it will be noted that a linear relationship is recovered for ${ Pr}_{M}\le 0.1$ }
      \label{fig:mouse}
   \end{figure}

  It is clear that the difference-curves spends its time in all the four quadrants. In the first quadrant, it would appear that the presence of pressure causes difference in the vorticity equations to grow momentarily faster than a comparative growth in the magnetic difference. But the growth in the magnetic difference soon supersedes that of the vorticity and we see a turning, wherein the magnetic difference continues to grow while the vorticity difference declines. The difference in the vorticity return to zero, where the effect of pressure modifies the growth such that there appears to be no difference with the case of pressure-less The difference then enters the fourth quadrant where magnetic difference reaches a maximum and begins to decline, while the vorticity difference is now increasingly negative. The magnetic difference then goes to zero where equality is achieved between the pressure-less and pressured equations. From this point on the difference enter the third quadrant where both differences are now both increasingly negative, meaning that the strength of the fields are swapped in both cases. It is conceivable that the graph of such difference for a true analogue would spend its life in the third quadrant. This amounts to eliminating or narrowing the crossover region in Figs. (\ref{fig:gull1}) and (\ref{fig:tiger1}). One way of doing this is to scan a parameter space made of ${Pr}_{M}\neq1$. The result of this is given in Fig.(\ref{fig:mouse}), where it is found that analogue status can be achieved for sufficiently small Prandtl numbers. It is known that ${Pr}_{M}\ll1$ means that thermal diffusivity dominates the flow as opposed to ${Pr}_{M}\gg1$ where momentum diffusivity would ordinarily dominate. The importance of the magnetic Prandtl number has also been studied in the context of MHD turbulence in \cite{Sheb} and \cite{Pon}
\begin{widetext}
\end{widetext}

\section{Conclusion}
In this article we have presented an in depth study of the magnetic and vorticity induction equations. We have defined a new parameter ${S}_{M}$ that we call the {\it Source number} which is given by the ratio of the battery term to the diffusive term, and which will be useful in classifying flows that include source terms. We have analyzed the evolutionary behaviors of the two fields with and without the presence of pressure, with the aim of finding constraints that would allow the two to be analogues. We find that, besides the case where the two equations are analogues when dissipation is neglected in the pressure-free case, analogy is still be achievable via suitable variation of Prandtl numbers for cases where pressure is non negligible. This in our view, strengthens the case for a theory of magnetic fluid analogues. 
\section{Acknowledgement}
BO acknowledges URC funding support administered by the University of Cape Town. PA acknowledges funding support from the National Research Foundation (NRF) of South Africa, as well as funding from the University of Cape Town, both administered by the Postgraduate Funding Office (PGFO) of the University of Cape Town. Both authors are grateful to the reviewers for the valuable comments.

\section{Appendix}

\appendix
\section{\label{appBat}Source Number}
In this section, we define a new dimensionless parameter denoted by ${S}_{M}$, which we call the battery number. It is given by the ratio of the battery term to the diffusion term. The battery term in its original form is $c ( k_{b}\nabla n_{e}\times\nabla T)/e n_{e}$, where $c$ is the velocity of light ( dimension $LT^{-1}$, $k_{b}$ the Boltzmann constant ( dimension $L^{2} M T^{-2} \Theta^{-1}$), $n_{e}$ is the electron number density (dimension; $L^{-3}$), $T$  is the temperature of the plasma fluid (dimension $\Theta$), and $e $ is the charge (dimension; $IL^{-1}$) . Writing the variable and constants that make the battery terms in terms of dimensional units, one will find: $ML^{2}T^{-3}I^{-1}$, which is the same units found when one write the diffusive term in terms of its dimensions. This means that ${S}_{M}$ is a dimensionless quantity.


\end{document}